\documentclass[%
 reprint,
nobibnotes,
 amsmath,amssymb,
 aps,
 prl,
 longbibliography
]{revtex4-2}

\usepackage{graphicx}
\usepackage{dcolumn}
\usepackage{bm}

\usepackage{xcolor}

\begin{document}

\preprint{APS/123-QED}

\title{Out-of-equilibrium selection pressure enhances inference from protein sequence data}
\author{Nicola Dietler} 
\author{Cyril Malbranke} 
\author{Anne-Florence Bitbol}
\email{anne-florence.bitbol@epfl.ch}
\affiliation{Institute of Bioengineering, School of Life Sciences, École Polytechnique Fédérale de Lausanne (EPFL), CH-1015 Lausanne, Switzerland}
\affiliation{SIB Swiss Institute of Bioinformatics, CH-1015 Lausanne, Switzerland}
\date{\today}

\begin{abstract}
Homologous proteins have similar three-dimensional structures and biological functions that shape their sequences. The resulting coevolution-driven correlations underlie methods from Potts models to AlphaFold, which infer protein structure and function from sequences. Using a minimal model, we show that fluctuating selection strength and the onset of new selection pressures improve coevolution-based inference of structural contacts. Our conclusions extend to realistic synthetic data and to the inference of interaction partners. Out-of-equilibrium noise arising from ubiquitous variations in natural selection thus enhances, rather than hinders, the success of inference from protein sequences. 
\end{abstract}

\maketitle

\paragraph*{Introduction.}
During evolution, proteins evolve through random mutations of their sequences and natural selection for function. This coupling between sequence and function leads to coevolution between amino acid sites in multiple sequence alignments (MSAs) of homologous proteins, which share ancestry, three-dimensional structure, and function. Amino-acid identities at contacting positions in a protein's three-dimensional structure are correlated, due to the need to preserve physico-chemical complementarity. This has allowed to predict structural contacts from sequences, using mutual information~\cite{Dunn08} and Potts models~\cite{Weigt09,Morcos11,Marks11,Barton16}, which are pairwise maximum entropy models inferred on MSAs. Attention coefficients in protein language models trained on MSAs capture coevolution~\cite{rao2021msa,Jumper21}, a key factor underlying AlphaFold's success in protein structure prediction~\cite{Jumper21}. Beyond structure, coevolution allows to infer interaction partners~\cite{Burger08,Bitbol16,Bitbol18,Lupo24} and functional sectors of collectively correlated amino acids~\cite{Halabi09,Rivoire16,Wang19,Dietler24}, as well as to investigate protein evolution~\cite{delaPaz20,DiBari24,DeLeonardis25,biswas_kinetic_2024,gizzio_evolutionary_2024,alvarez_vivo_2024}. 

Natural environments often change over time~\cite{Mustonen2008,Bell2010,Melbinger15,Wienand17,Taitelbaum20,Meyer20}, causing variable selection pressures on proteins~\cite{Mustonen2008}.
How do such fluctuations impact inference from sequences?
To address this, we consider a minimal model of protein sequences evolving under fluctuating selection to preserve structure. 
We show that the resulting out-of-equilibrium noise can enhance the inference of structural contacts via Potts models. This conclusion extends to realistic synthetic data. It also holds in a simple model of the onset of a new selection pressure, where a random ancestral sequence evolves under structural selection with phylogeny. Finally, we show that out-of-equilibrium selection also enhances the inference of interaction partners. Our results suggest that time-dependent external drives can facilitate inference by pairwise maximum entropy models.

\paragraph*{Model and methods.}
We consider a minimal model, where protein sequences are represented as length-$L$ sequences of  Ising spins $(\sigma_1,\dots,\sigma_L)$, defined on the nodes of a fixed Erd\H{o}s-Rényi random graph. Structural contacts are modeled by pairwise ferromagnetic couplings on the graph edges, all set to 1 for simplicity, leading to the Hamiltonian 
\begin{equation}
H=-\sum_{(i,j)\in\mathcal{E}}\sigma_i\sigma_j\,,   
\label{ham}
\end{equation}
where $\mathcal{E}$ denotes the set of graph edges, see also~\cite{Dietler2023}. 
We sample independent equilibrium sequences using a Metropolis–Hastings algorithm by proposing spin flips at random sites and accepting them with probability 
\begin{equation}
p=\min\left[1,\exp(-\Delta H/T)\right]\,,
\label{proba}
\end{equation}
where $\Delta H$ represents the energy variation associated to the flip, and $T$ the Monte Carlo sampling temperature, see Fig.~S1(a) for a schematic and Fig.~S2 for information about equilibration. 
Selection strength is controlled by $T$, with higher $T$ corresponding to weaker selection~\cite{Dietler2023}. 
We model variable selection strength using a telegraph process that switches between $T_1$ and $T_2>T_{1}$ with equal transition rates $1/\tau$. We start from independent equilibrium sequences at temperature $T_{1}$, see Fig.~S1(b). 

To go beyond this minimal model, we generate realistic synthetic data using a Potts model inferred from natural sequences via bmDCA~\cite{Figliuzzi18,Russ20}, an approach whose generative power was experimentally validated~\cite{Russ20}. 

Finally, to model the onset of a new selection pressure, we start from a random ancestral sequence and apply selection for structure using the minimal model described above. We evolve different sequences from the same ancestor via independent Markov chains, producing a simple star phylogeny, see Fig.~S1(c). 

\paragraph*{Fluctuating selection enhances contact inference.} In our model, selection strength is controlled by the sampling temperature $T$. At very low $T$, only spin flips that decrease the energy (Eq.~\ref{ham}) are accepted, leading to low sequence variability and poor inference. Conversely, at very high $T$, almost all spin flips are accepted, producing noisy sequences that also hinder inference. As a result, inference is optimal at intermediate temperatures. In the present minimal model featuring a ferromagnetic-paramagnetic phase transition (Fig.~S3 and~\cite{Gandarilla20,Dietler2023}), inference is most successful around and moderately above this transition~\cite{Ngampruetikorn22,Dietler2023}. With our parameter values, it occurs at $T_C=4$ (Fig.~S4). Inference performance also depends on the number of available sequences (Fig.~S5 and~\cite{Gandarilla20,Ngampruetikorn22}). In natural data, this number is limited by experimental sampling and by biological evolution, leading to finite-size effects that hinder inference~\cite{Morcos11,Marks11,Ovchinnikov14}. Thus motivated, we work with finite MSAs throughout.

How does the time variability in selection pressure impact inference from sequence data? To answer this within our minimal model, we start from independent sequences equilibrated at $T_{1} = 1$ and evolve them under a time-varying sampling temperature that switches between $T_1 = 1$ and $T_2 = 15$. 

\begin{figure}[htb!]
\centering
\includegraphics[width=\columnwidth]{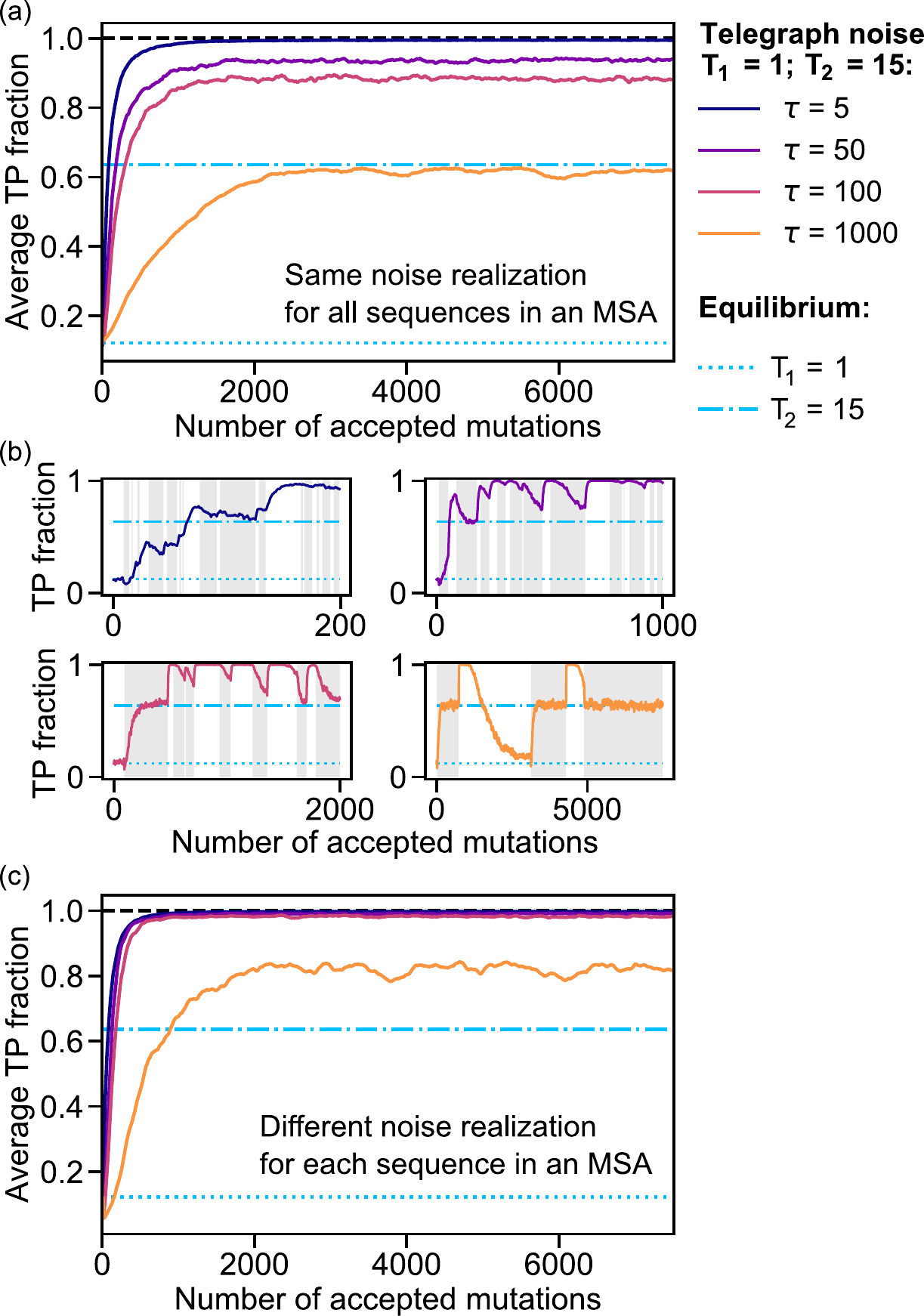}
\caption{\textbf{Impact of fluctuating selection strength on contact prediction.} TP fraction versus number of accepted mutations per sequence, while switching selection strength (i.e.\ the sampling temperature) via a telegraph process with timescale $\tau$. (a) For each value of $\tau$, the TP fraction is averaged over 1000 replicates. In each replicate, the same telegraph process is used for all sequences in the MSA. Different replicates use different realizations of the data generation and of the telegraph process. (b) The TP fraction is shown for single realizations of the telegraph process and data generation. Gray background: $T_{2}$; white background: $T_{1}$. (c) Same as in (a), except that a different telegraph process is used for each sequence in an MSA. In all panels, 
TP fractions for equilibrium sequences generated at $T_1$ and $T_2$ are shown for reference. 
Sequences are generated using our minimal model, see Fig.~S1(a-b), with $T_{1} = 1$ and $T_{2} = 15$. For each realization, we generate an MSA of 2048 sequences of length 200 using an Erd\H{o}s-Rényi random graph with probability 0.02 to represent contacts, and infer contacts via mean-field DCA (mfDCA)~\cite{Marks11,Morcos11}, which is computationally efficient, and performs well on minimal data~\cite{Dietler2023}, with pseudocount $0.01$. 
\label{fig:tpfrac_telegraph}}
\end{figure}

Fig.~\ref{fig:tpfrac_telegraph} shows the True Positive (TP) fraction -- defined as the fraction of true contacts among the top $N$ inferred Ising couplings, where $N$ is the actual number of contacts -- versus the number of accepted spin flips, representing mutations. Panel (a) shows that the average TP fraction increases with time before plateauing, indicating that fluctuating selection strength enhances inference performance. Faster temperature switches further improve inference by increasing sequence diversity. The effective number of sequences, $M_\mathrm{eff}$~\cite{Weigt09}, defined by grouping sequences with Hamming distance smaller than 0.2, reaches $1980$. This is close to the value for an MSA generated at $T=15$ ($2048$), and to the total number of sequences per MSA ($2048$), and far above the nearly frozen $M_\mathrm{eff}=2$ at $T=1$. Inference under variable temperature outperforms both fixed-temperature cases at $T_1$ and $T_2$, and matches the best fixed-temperature case, see Fig.~S4, demonstrating that variable selection strength can substantially improve inference.

Fig.~\ref{fig:tpfrac_telegraph}(b) shows that, in single realizations of the process, switching to strong selection ($T_1$) after a weak selection phase ($T_2$) yields a transient improvement in inference performance (see also Fig.~S6). Increasing selection strength suppresses noise that partly obscures structural correlations, before significantly reducing sequence diversity. Thus, switching between weak and strong selection temporarily enhances structural correlations in sequences. If the system is held at a fixed temperature for times comparable to the equilibration times (see Fig.~S2), inference performance relaxes to its equilibrium value at that temperature. The observed improvement is therefore a transient, out-of-equilibrium effect. Faster temperature alternations, on timescales significantly below equilibration, amplify these effects and further enhance inference performance, as seen in Fig.~\ref{fig:tpfrac_telegraph}(a).  

What is the impact of the temperatures $T_1$ and $T_2$ between which the system switches? In our minimal model, inference is optimal around and slightly above the ferromagnetic-paramagnetic transition temperature $T_C$ (see above). Accordingly, alternating between $T_1<T_C$ and $T_2>T_C$ improves inference, whereas switching between two temperatures below or above $T_C$ does not (Fig.~S7). Furthermore, for fast switches, the data resembles equilibrium data generated at an intermediate temperature between $T_1$ and $T_2$, and leads to similar inference performance (Fig.~S8). 

So far, all sequences in an MSA were subjected to the same fluctuating selection strength. However, in natural data, different sequences (e.g.\ from different species) may experience different selection histories. Fig.~\ref{fig:tpfrac_telegraph}(c) shows inference performance when each sequence in an MSA is subjected to an independent realization of the temperature fluctuations. Inference is enhanced even more strongly than in Fig.~\ref{fig:tpfrac_telegraph}(a), indicating that heterogeneous fluctuating selection further amplifies the effect.

\paragraph*{Extension to realistic synthetic data.}
How do the observations made with our minimal model generalize to more realistic data? To address this question, we infer a Potts model from a natural MSA of the protein family PF0004 (AAA ATPase) from Interpro/Pfam~\cite{mistry2021pfam} via bmDCA~\cite{Figliuzzi18,Russ20}. We then generate sequences from this Potts model, either at equilibrium or under fluctuating selection strength (see Model and Methods). 

\begin{figure}[htb!]
\centering
\includegraphics[width=0.9\columnwidth]{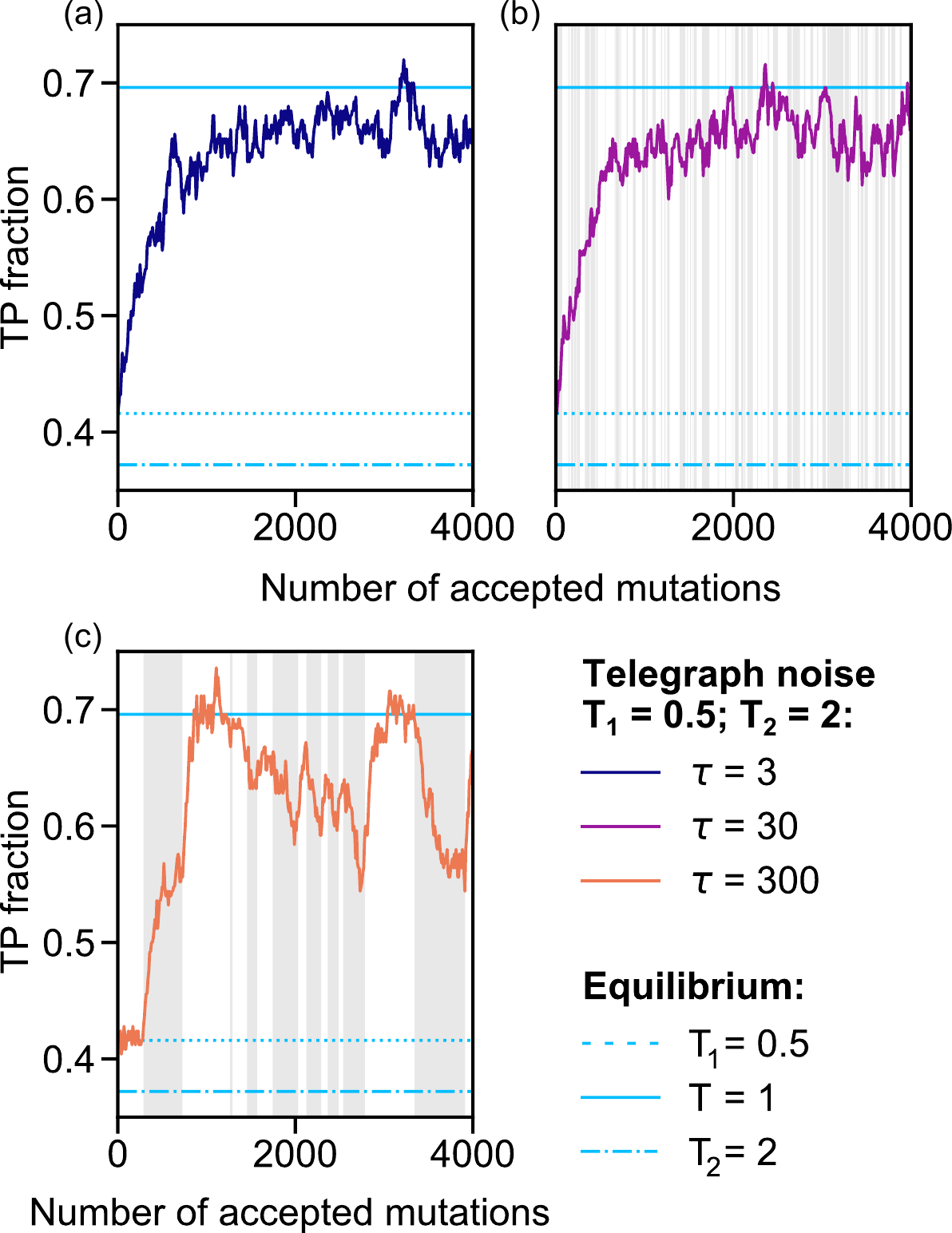}
\caption{\textbf{Impact of fluctuating selection strength on contact prediction: PF0004 family.} 
As in Fig.~\ref{fig:tpfrac_telegraph}(b), the TP fraction is shown versus the number of accepted mutations under switching selection strength via a telegraph process, but for realistic data composed of MSAs of 70,000 sequences generated from a Potts model inferred on a natural MSA of 39,277 sequences from the PF0004 family (with length 132), following Refs.~\cite{Lupo22,Figliuzzi18}, in particular using regularization strengths 0.01. 
Gray background: $T_{2}$; white background: $T_{1}$ (alternations not shown in the first panel for readability). Inference performances for equilibrium sequences generated at $T_1$, $T_2$, and $T=1$, are shown for reference. 
Contact inference is performed using plmDCA~\cite{Ekeberg13,Ekeberg14}, with regularization strengths set to 0.01 and no phylogenetic reweighting.
\label{fig:tpfrac_telegraph_pf4}}
\end{figure}

Fig.~\ref{fig:tpfrac_telegraph_pf4} shows the TP fraction versus the number of accepted mutations for this realistic synthetic data. As in the minimal model, alternating between strong and weak selection improves inference performance, with faster alternations between low and high temperature yielding larger gains. Inference performance then becomes comparable to the best equilibrium performance, obtained at temperature $T = 1$ (Fig.~S9), even though both alternating temperatures yield poorer inference at equilibrium. Thus, our conclusions from the minimal model extend to realistic synthetic data. 

\paragraph*{Impact of selection onset with phylogeny.}
Beyond fluctuating environments, out-of-equilibrium effects also arise when a new selection pressure emerges. 
We model this by generating sequences starting from a random ancestral sequence, under selection for structural contacts at fixed $T$ in the minimal model, using a simple star phylogeny (see Model and Methods). This setup captures the onset of a specific selection pressure acting on a protein (sub)family, starting from an ancestor that was not subject to this particular constraint. 

\begin{figure}[htb!]
\centering
\includegraphics[width=0.9\columnwidth]{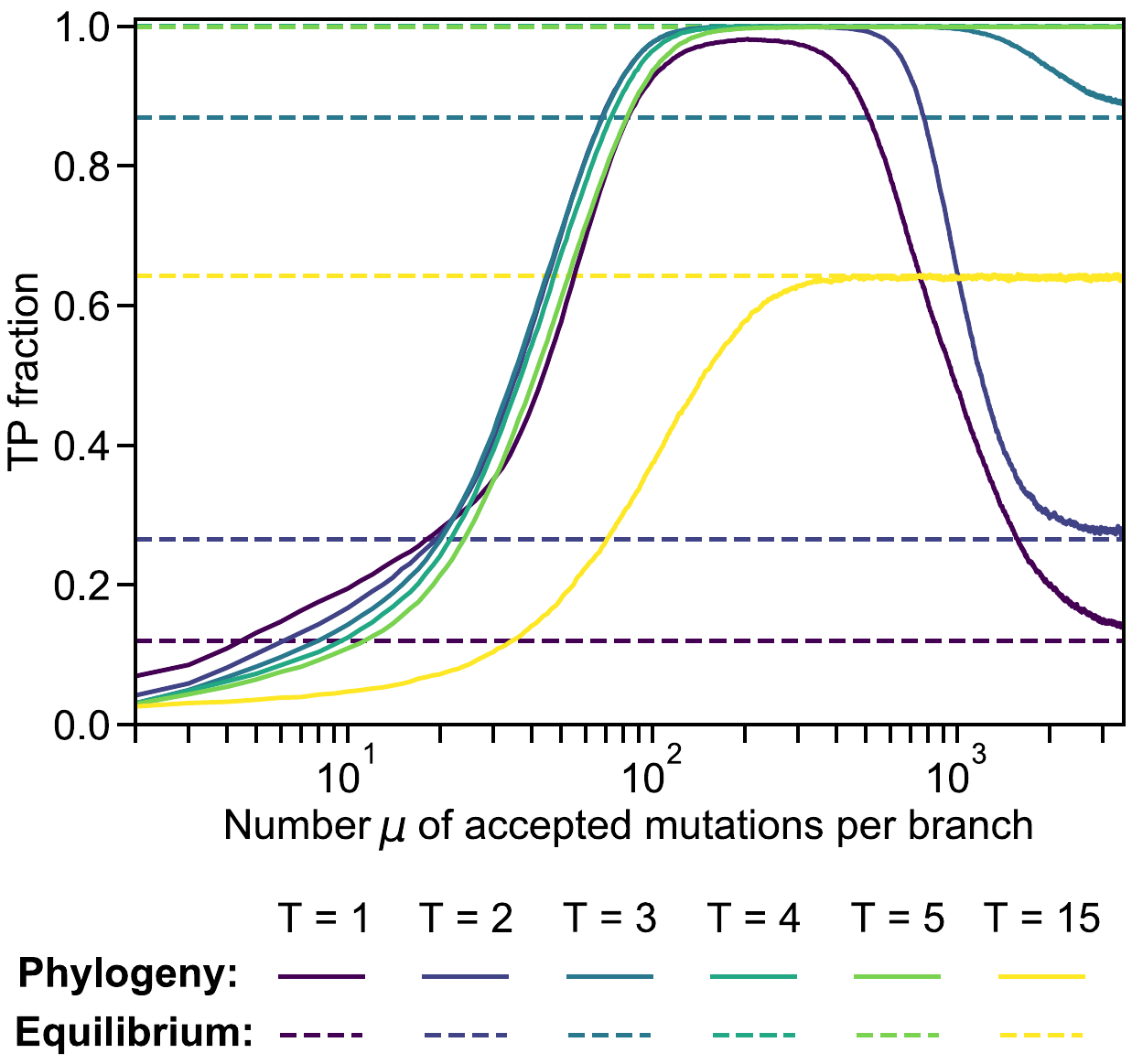}
\caption{\textbf{Impact of selection onset on contact inference with phylogeny.} The TP fraction is shown versus the number $\mu$ of accepted mutations per branch of the star phylogenetic tree, starting from a random ancestral sequence. Sequences with length 200 are generated as explained in Fig.~S1(c), using the same Erd\H{o}s-Rényi graph as in Fig.~\ref{fig:tpfrac_telegraph}, along a star phylogenetic tree with 2048 branches. Along each branch, Monte Carlo sampling is performed until $\mu$ mutations are accepted, giving 2048 final sequences, which constitute our MSA. We infer contacts via mfDCA~\cite{Marks11,Morcos11,Dietler2023} with pseudocount 0.01. Results are averaged over 100 replicates, each starting from a different random ancestor.}
\label{fig:tpfrac_vs_mu}
\end{figure}

Fig.~\ref{fig:tpfrac_vs_mu} shows the TP fraction versus the number $\mu$ of accepted mutations per branch in the star phylogeny. For $T<4$, i.e.\ below the ferromagnetic-paramagnetic transition of the minimal model, the TP fraction exhibits a maximum at intermediate $\mu$. Inference then performs significantly better than the equilibrium TP fraction approached at large $\mu$. This enhancement coincides with a transient increase of sequence diversity (Fig.~S10), and persists for a balanced binary branching tree (Fig.~S11).  
Hence, under strong selection, out-of-equilibrium noise originating from the random ancestor enhances contact inference. The beneficial effect of out-of-equilibrium noise evidenced for fluctuating selection thus also applies to the onset of a new selection pressure, and remains effective even with phylogeny, which impairs contact inference performance under constant selection~\cite{Dietler2023}. 

\begin{figure}[htb!]
\centering
\includegraphics[width=\columnwidth]{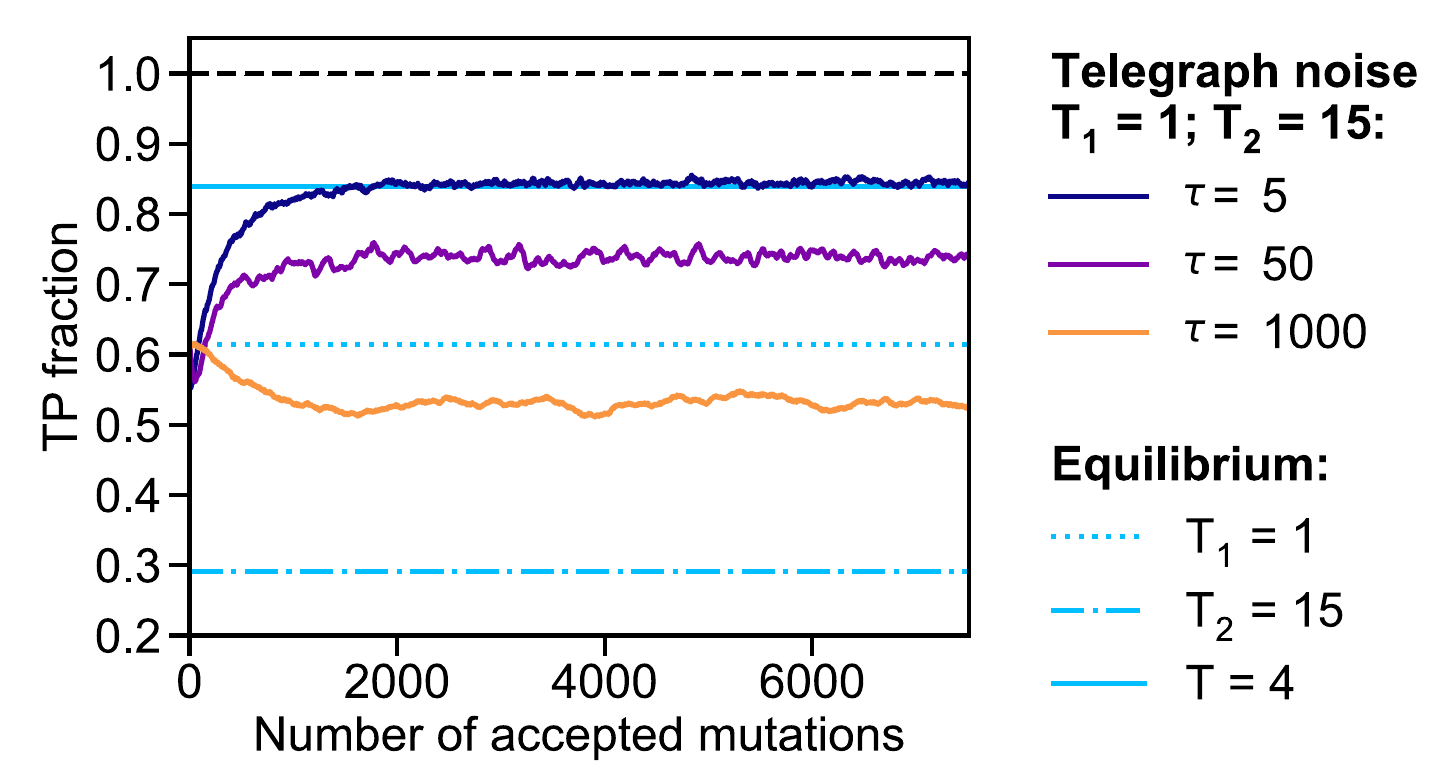}
\caption{\textbf{Impact of fluctuating selection strength on interaction partner prediction.} TP fraction versus number of accepted mutations, while switching selection strength (i.e.\ the sampling temperature) via a telegraph process with timescale $\tau$. For each value of $\tau$, the TP fraction is averaged over 1000 replicates. TP fractions for equilibrium sequences generated at $T_1$ and $T_2$ are shown for reference. MSAs comprising 1024 sequences are generated as in Fig.~\ref{fig:tpfrac_telegraph}(a), and then randomly split into a training set of 400 sequences and a testing set of 624 sequences. The latter is randomly divided in sets of 4 sequences each, representing species, and each sequence is split in two halves of equal length representing two interaction partners. Pairings between partners are blinded in each species, and predicted using mfDCA scores with pseudocount $0.01$~\cite{Gerardos22}.}
\label{fig:tpfrac_partners}
\end{figure}

\paragraph*{Extension to interaction partner inference.}
So far, we considered structural contact inference. Coevolution-based methods are also used to infer protein interaction partners~\cite{Burger08,Bitbol16,Bitbol18,Gerardos22}. To test whether our results extend to this problem, we perform partner inference within our minimal model using the method of~\cite{Gerardos22}. As for contact inference, partner inference performance peaks at an intermediate temperature (Fig.~S4). Fig.~\ref{fig:tpfrac_partners} shows that fluctuating selection strength enhances partner inference, indicating that the beneficial effect of out-of-equilibrium selection extends beyond contact inference.

\paragraph*{Discussion.}
Using a minimal model, we showed that alternating between weak and strong selection enhances structural contact inference from sequences. This effect extends to realistic synthetic sequences. It also extends to the onset of a new selection pressure, even in the presence of phylogeny. Moreover, out-of-equilibrium selection also enhances interaction partner inference.

Under strong selection, limited sequence variability hinders coevolution-based inference. Alternating with weak selection, or starting from an unselected ancestor transiently increases diversity, enhancing inference. Overall, our results reveal that, far from hindering it, out-of-equilibrium selection can facilitate the extraction of structural and functional information from protein sequences. While we relied on synthetic data, which allowed us to vary selection pressures and control ancestral sequences, the extension to realistic data suggests that these principles are generic and relevant for natural proteins, where variable selection pressures are ubiquitous~\cite{Mustonen2008,Bell2010}. Hence, our findings suggest that the success of coevolution-based inference in proteins may be partly caused by fluctuating selection intensity. 

We considered simple forms of selection variability, either switching between two strengths or adding a new selection pressure. Other forms of fluctuations of selection strengths, including general random walks, are expected to yield similar results, provided that weak and strong selection are included and the fluctuations are faster than equilibration times. Besides, while we focused on structural selection pressures, which are a major constraint on protein evolution~\cite{morcos_coevolutionary_2014}, our work could be extended to other evolutionary constraints, such as those shaping collectively correlated amino acid groups~\cite{Halabi09,Rivoire16,Wang19,Dietler24}, and to RNA structure inference~\cite{Zerihun19}. Additional perspectives include selection pressure variations that cannot be captured by an overall scaling of selection strength, e.g.\ affecting specific fields or couplings. 
Finally, although we focused on inference based on Potts models, coevolution also underlies recent deep learning approaches, including AlphaFold~\cite{Jumper21}, indicating that out-of-equilibrium fluctuations may similarly benefit these approaches. 

Beyond natural evolution, our results have implications for directed evolution experiments~\cite{Arnold18}. 
In recent studies~\cite{Fantini2019,STIFFLER2020}, structural prediction was performed using sequences evolved \textit{in vitro} under constant selection strength. It was later shown~\cite{Bisardi2021} that tuning this strength can improve contact inference.
Our results suggest that dynamically varying selection strength during experiments, e.g.\ by varying antibiotic concentration, could also enhance inference.

More broadly, pairwise maximum entropy models are widely used for inference beyond protein sequence data. A prominent example is neuronal population activity~\cite{Schneidman06,meshulam_collective_2017,meshulam_statistical_2025}. Neurons are often driven by common input stimuli that vary in time, and could play a role analogous to fluctuating selection strength in proteins. Such inputs have been found to impact inference~\cite{tyrcha_effect_2013}, increase inferred couplings in some cases~\cite{roudi_ising_2009}, and generate signatures reminiscent of criticality~\cite{schwab_zipfs_2014,ngampruetikorn_extrinsic_2025}, which are frequently observed in biological systems~\cite{Mora11}. Our results thus suggest a potentially general principle: time-dependent external drives can enhance inference in systems described by pairwise maximum entropy models.

\paragraph*{Code accessibility.} Our code is available at \url{https://github.com/Bitbol-Lab/OutOfEquilibrium-ProtSeq}.

\begin{acknowledgments}
This research was partly funded by the European Research Council (ERC) under the European Union’s Horizon 2020 research and innovation programme (grant agreement No.~851173, to A.-F.~B.).
\end{acknowledgments}




%

\onecolumngrid

\renewcommand{\thefigure}{S\arabic{figure}}
\setcounter{figure}{0}
\renewcommand{\thetable}{S\arabic{table}}
\setcounter{table}{0} 
\renewcommand{\theequation}{S\arabic{equation}}
\setcounter{equation}{0}
\preprint{APS/123-QED}

\newpage

\phantom{SI}
\vspace{1cm}

\begin{center}
	\huge{\textbf{Supplementary material}}
\end{center}

\vspace{3cm}

\begin{figure}[htb!]
	\centering
	\includegraphics[width=0.85\textwidth]{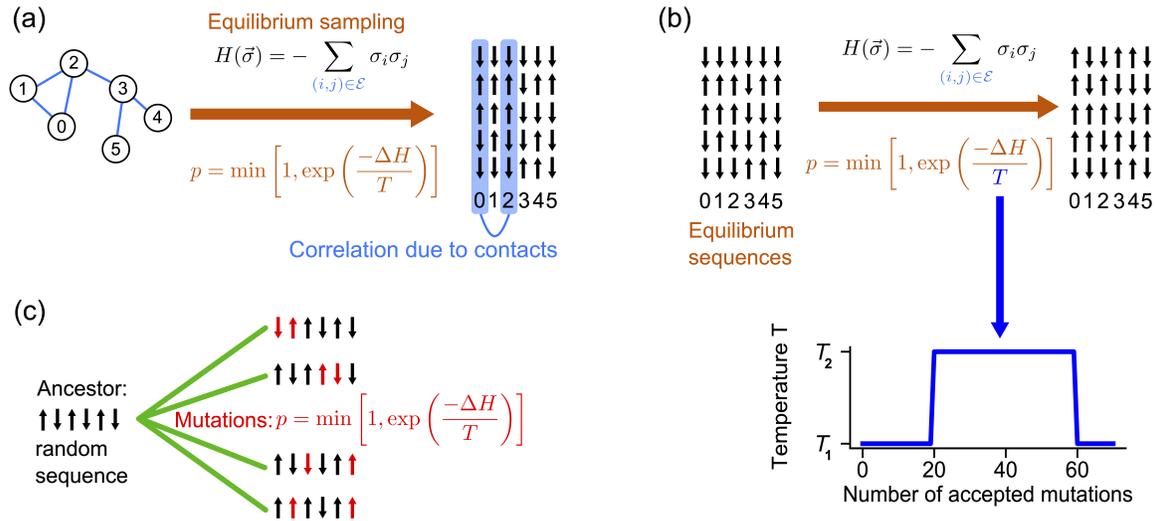}
	\caption{\textbf{Minimal model}. (a) We model structural contacts as couplings set to 1 on the edges of an Erd\H{o}s-Rényi random graph. Couplings between other nodes are set to 0. 
		Equilibrium sampling of independent sequences is performed using a Metropolis--Hastings algorithm under the Hamiltonian in Eq.~(1), with spin flip acceptance probability in Eq.~(2), starting from random initial sequences. The MSA formed from the resulting sequences features correlations arising from couplings, shown between columns $0$ and $2$ (blue). (b) To generate sequences with fluctuating selection strength, we start from sequences generated at equilibrium with temperature $T_{1}$. We then evolve each of these sequences using the same Metropolis--Hastings algorithm as before, but with a sampling temperature $T$ that switches between $T_{1}$ and $T_{2}$. (c) To minimally model the impact of emerging selection during evolution along a phylogeny, we take a random sequence as ancestor, and we evolve it along a star phylogenetic tree (green) where mutations (red) are accepted with the probability in Eq.~(2). This represents selection for structural contacts being switched on. On each branch of the tree, $\mu$ mutations are accepted, with $\mu=2$ in our schematic.}
	\label{fig:fig_conceptual_outofeq}
\end{figure}

\begin{figure}[htb!]
	\centering
	\includegraphics[width=0.85\columnwidth]{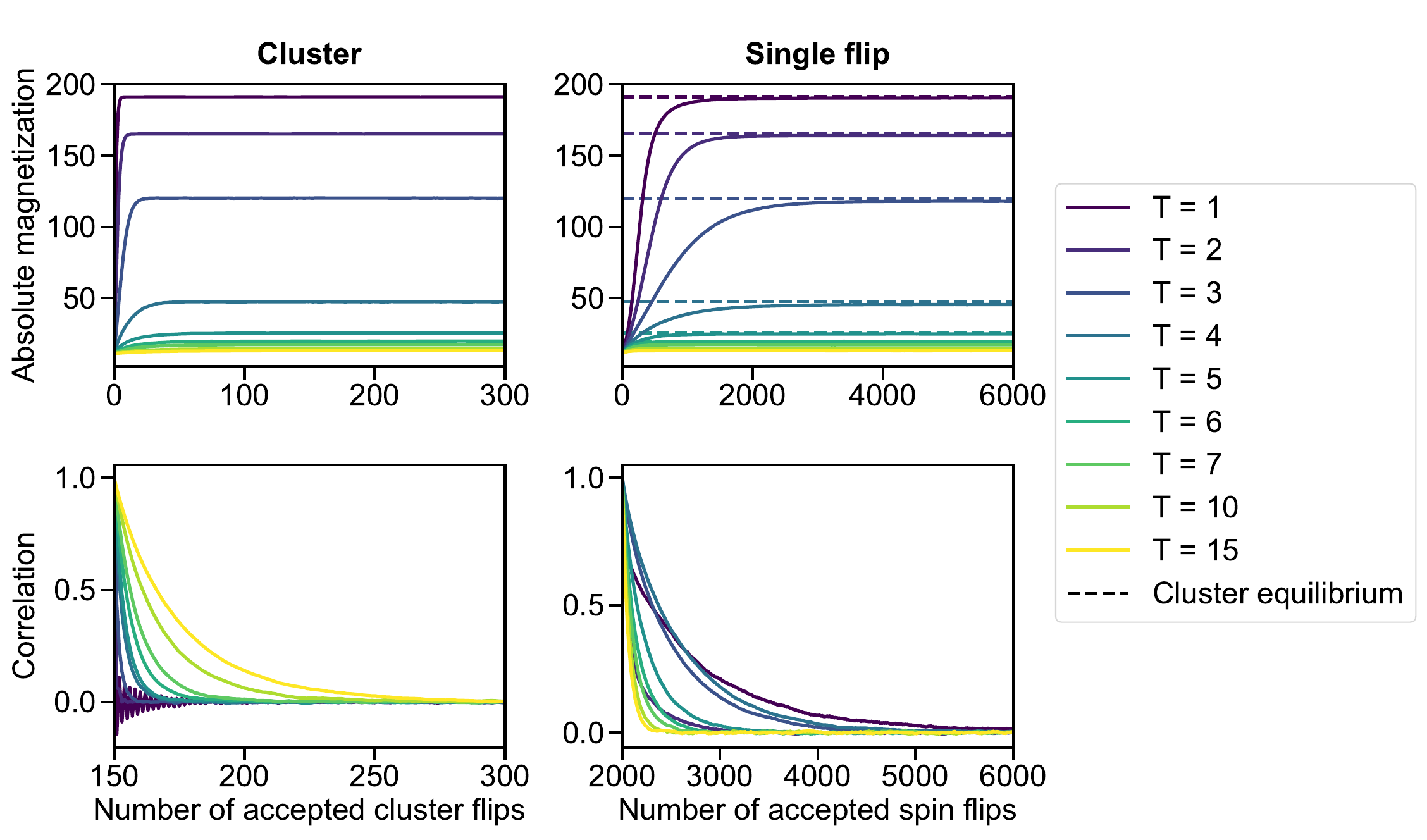}
	\caption{\textbf{Equilibration of sequences.} We report results regarding the generation of equilibrium sequences at fixed sampling temperature by two Metropolis Monte Carlo algorithms, namely the Wolff cluster algorithm~\cite{Krauth} (left panels) and the single flip algorithm (right panels). Note that, elsewhere in this work, the single flip algorithm is used if not stated otherwise, given its biological relevance (single flips representing mutations). We start from sequences where each spin is chosen uniformly at random. Absolute magnetization (top) and correlation of the absolute magnetization with that of the initial state (bottom) are shown versus the number of accepted cluster flips (left) or single spin flips (right). As equilibrium is reached, magnetisation plateaus and correlation tends to zero. This process is faster with the cluster algorithm. For comparison, the final value or magnetization reached with this algorithm is shown on the top right panel. At each sampling temperature, 2048 sequences are generated, each starting from a different random sequence. Results are averaged over 100 replicates. 
	}
	\label{fig:Corr_Magn_TP_CLSF}
\end{figure}

\begin{figure}[htb!]
	\centering
	\includegraphics[width=0.9\columnwidth]{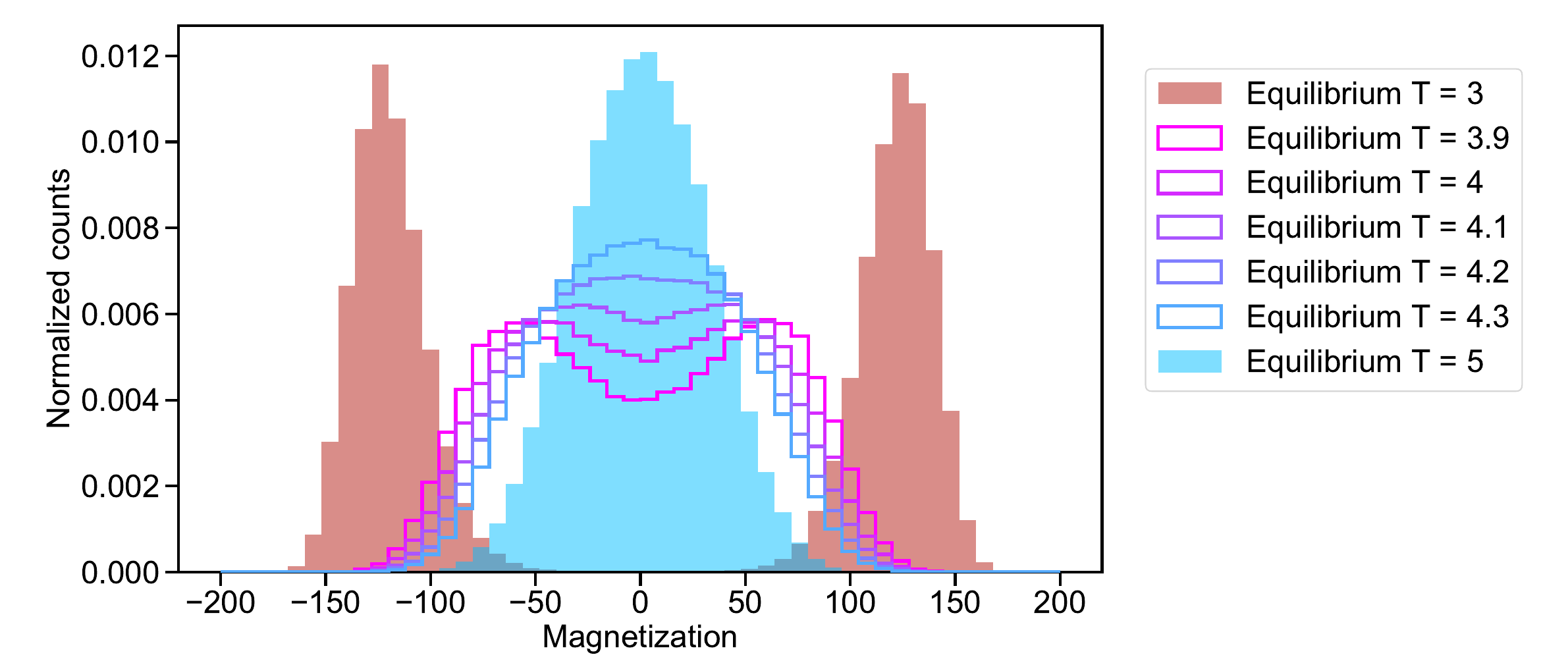}
	\caption{\textbf{Magnetization histograms evidencing a ferromagnetic-paramagnetic transition.} We report histograms of the magnetization at different sampling temperatures. Sequences are generated at equilibrium using the cluster algorithm, starting from random sequences and using 300 accepted cluster flips for each sequence. Each histogram is computed over 2048 sequences. The same Erd\H{o}s-Rényi graph as in Fig.~1 and elsewhere is used to represent contacts.}
	\label{fig:hist_Tc}
\end{figure}

\begin{figure}[htb!]
	\centering
	\includegraphics[width=0.9\columnwidth]{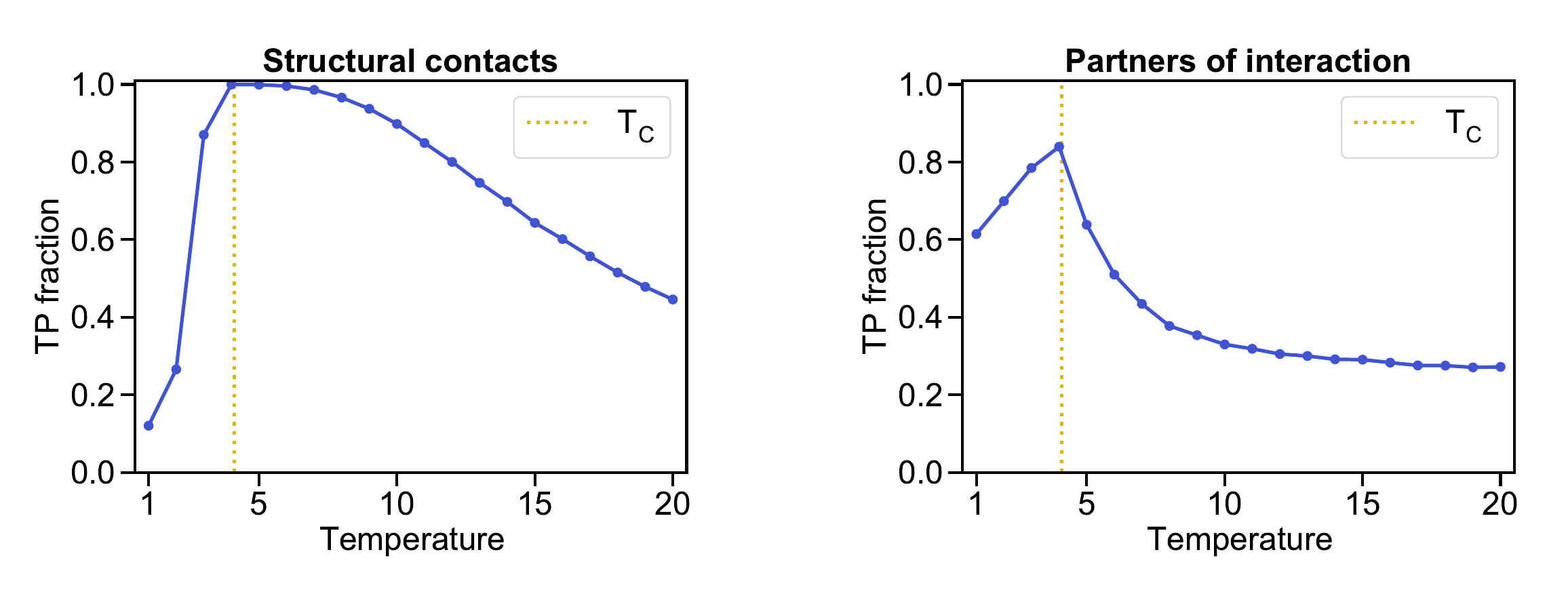}
	\caption{\textbf{Impact of sampling temperature on contact and partner predictions at equilibrium.} We show the TP fraction as a function of sampling temperature for contact prediction (left panel) and interaction partner prediction (right panel). Sequences are generated at equilibrium for different temperatures using the minimal model, see Fig.~\ref{fig:fig_conceptual_outofeq}(a-b), and the cluster algorithm (starting from random sequences and using 300 accepted cluster flips), see Fig.~\ref{fig:Corr_Magn_TP_CLSF}. For structural contact inference, each TP fraction is computed over 2048 sequences, using the same method as in Fig.~\ref{fig:fig_conceptual_outofeq}. For partner inference, each TP fraction is computed over 624 sequences, using the exact same method (and data split approach) as in Fig.~4. The approximate temperature $T_C$ of the ferromagnetic-paramagnetic transition, see Fig.~\ref{fig:hist_Tc}, is indicated by a vertical dotted line. All results are averaged over 100 replicates.}
	\label{fig:tpfrac_vs_T}
\end{figure}

\begin{figure}[htb!]
	\centering
	\includegraphics[width=0.7\columnwidth]{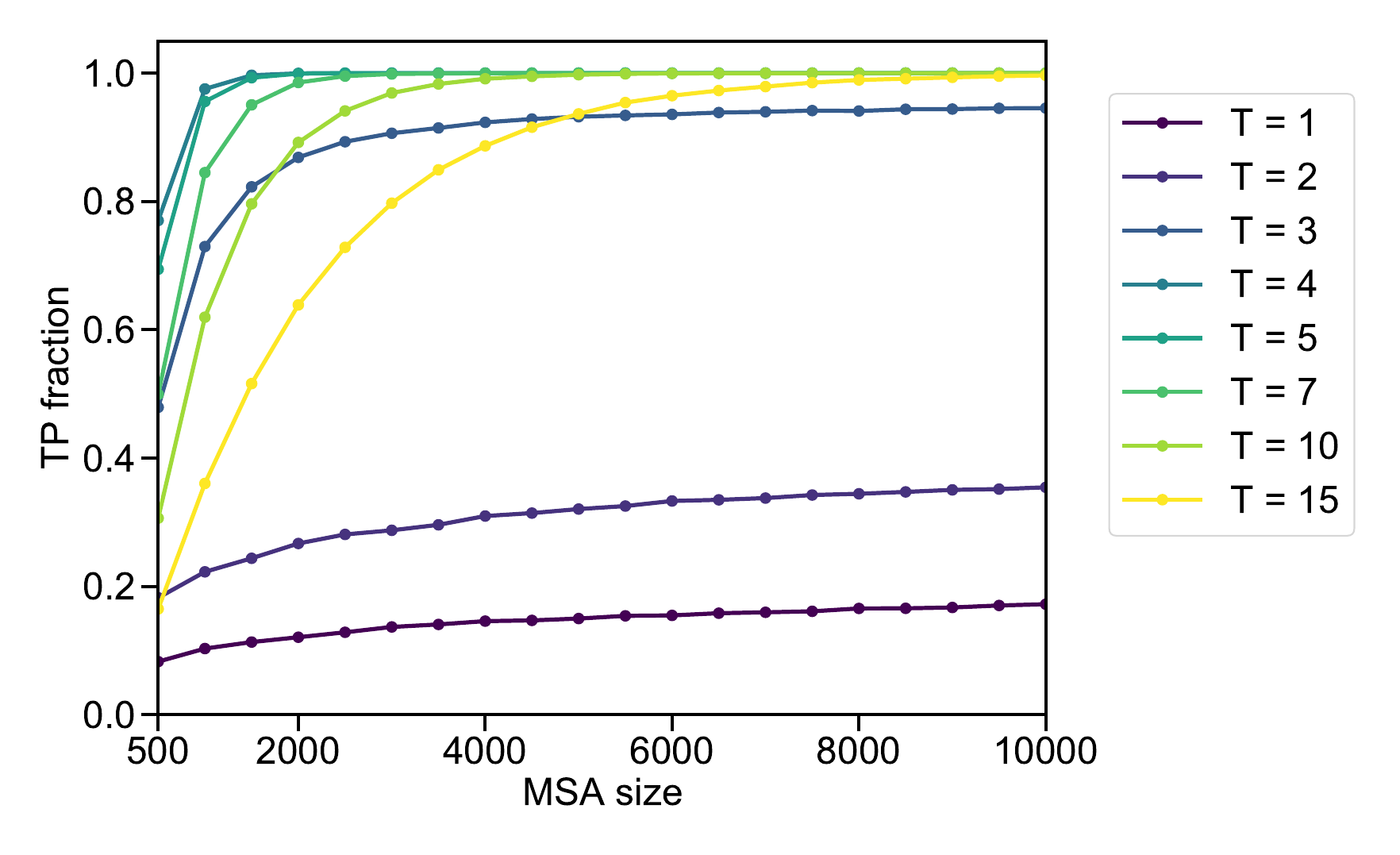}
	\caption{\textbf{Impact of MSA size on structural contact prediction.} We report the TP fraction for structural contact prediction versus the MSA size used to infer contacts. MSAs are generated at equilibrium for each sampling temperature considered (using the cluster algorithm, starting from random sequences and using 300 accepted cluster flips for each sequence). As elsewhere, we infer contacts via mfDCA~\cite{Marks11,Morcos11,Dietler2023} with pseudocount 0.01. }
	\label{fig:tp_msasize}
\end{figure}

\begin{figure}[htb!]
	\centering
	\includegraphics[width=0.9\columnwidth]{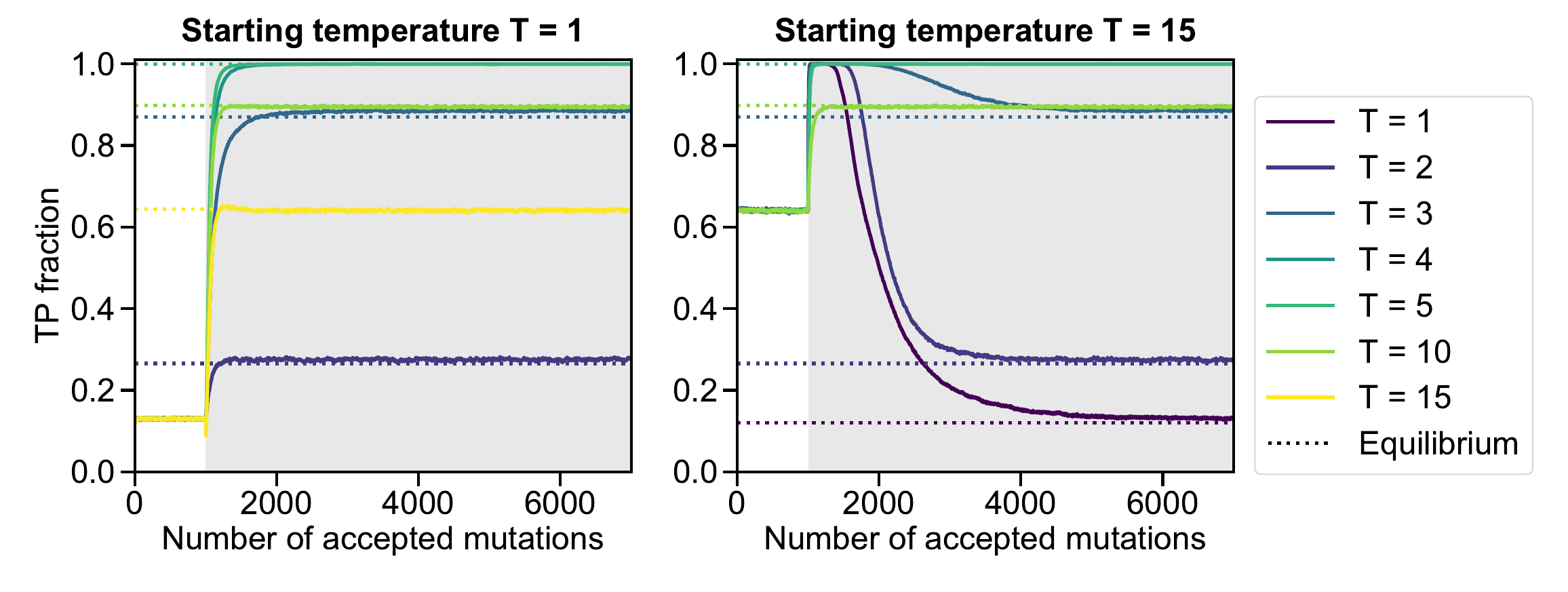}
	\caption{\textbf{Impact of switching the sampling temperature on the performance of contact inference. } We report the TP fraction for contact inference when starting from an equilibrated MSA at temperature $T = 1$ (left panel) or $T = 15$ (right panel) and switching to a different temperature. White background: initial temperature; grey background: new temperature. The initial MSA is generated with the cluster algorithm (starting from random sequences and using 300 accepted cluster flips) and then single flips are applied, 1000 being accepted at the initial temperature, before the switch occurs. In each case, the TP fraction is calculated using MSAs of 2048 sequences, and an averaging over 100 replicates is performed. 
	}
	\label{fig:quench_T1T2}
\end{figure}

\begin{figure}[htb!]
	\centering
	\includegraphics[width=0.75\columnwidth]{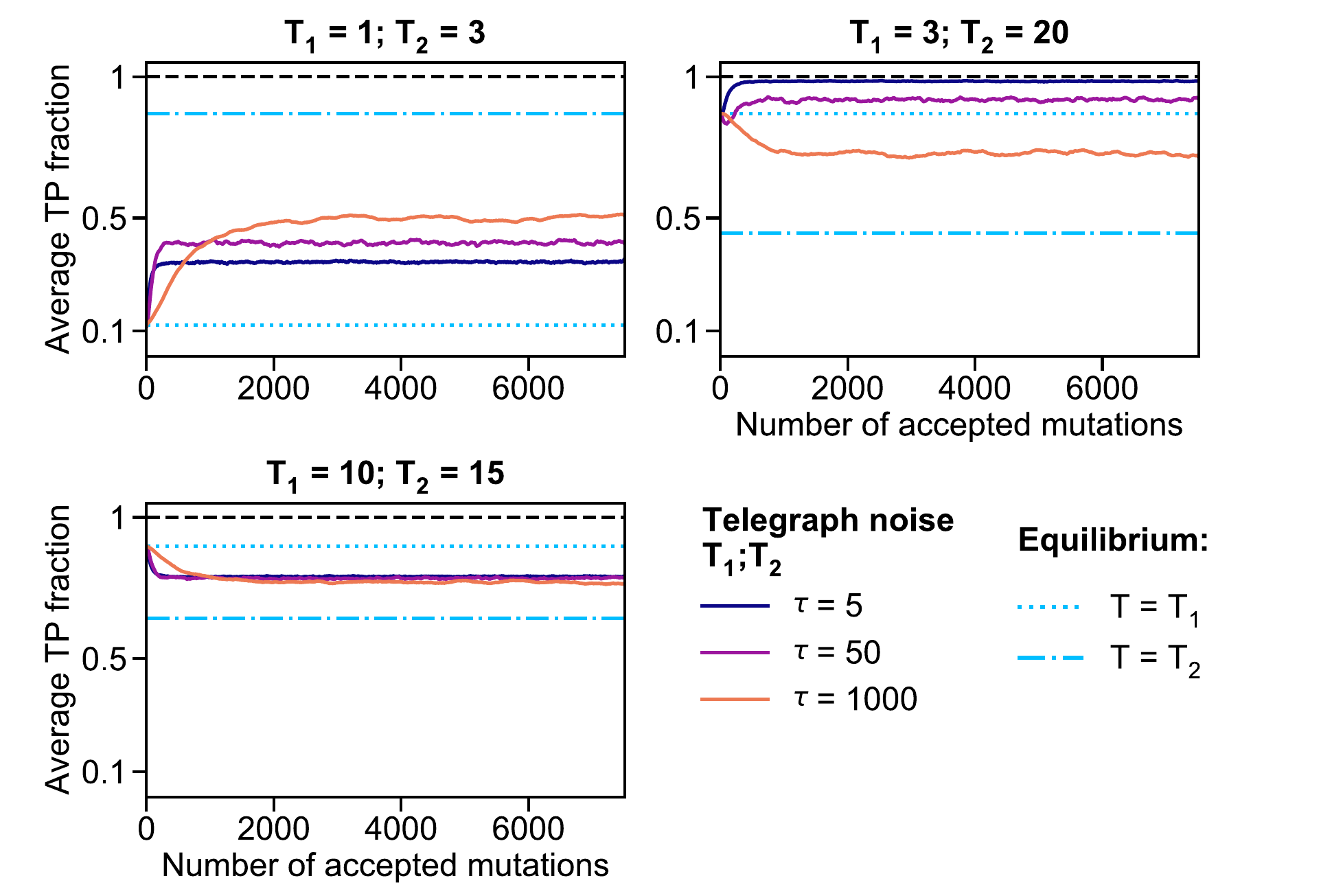}
	\caption{\textbf{Impact of fluctuating selection strength on contact prediction, using different sampling temperatures.} As in Fig~1(a), we report the TP fraction versus the number of accepted mutations, while switching selection strength (i.e.\ the sampling temperature) via a telegraph process with timescale $\tau$, but different pairs of temperatures $(T_1;T_2)$ are employed (one pair in each panel). For each value of $\tau$, the TP fraction is averaged over 1000 replicates. In each replicate, the same telegraph process is used for all sequences in the MSA. Different replicates use different realizations of the data generation and of the telegraph process. In all panels, TP fractions for equilibrium sequences generated at $T_1$ and $T_2$ are shown for reference. 
		Sequences are generated using our minimal model, see Fig.~\ref{fig:fig_conceptual_outofeq}(a-b). For each realization, we generate an MSA of 2048 sequences of length 200 using the same Erd\H{o}s-Rényi random graph with probability 0.02 to represent contacts as in Fig.~\ref{fig:fig_conceptual_outofeq}, and we infer contacts via mean-field DCA (mfDCA)~\cite{Marks11,Morcos11,Dietler2023}, with pseudocount $0.01$.}
	\label{fig:tpfrac_otherT1T2}
\end{figure}

\begin{figure}[htb!]
	\centering
	\includegraphics[width=0.95\columnwidth]{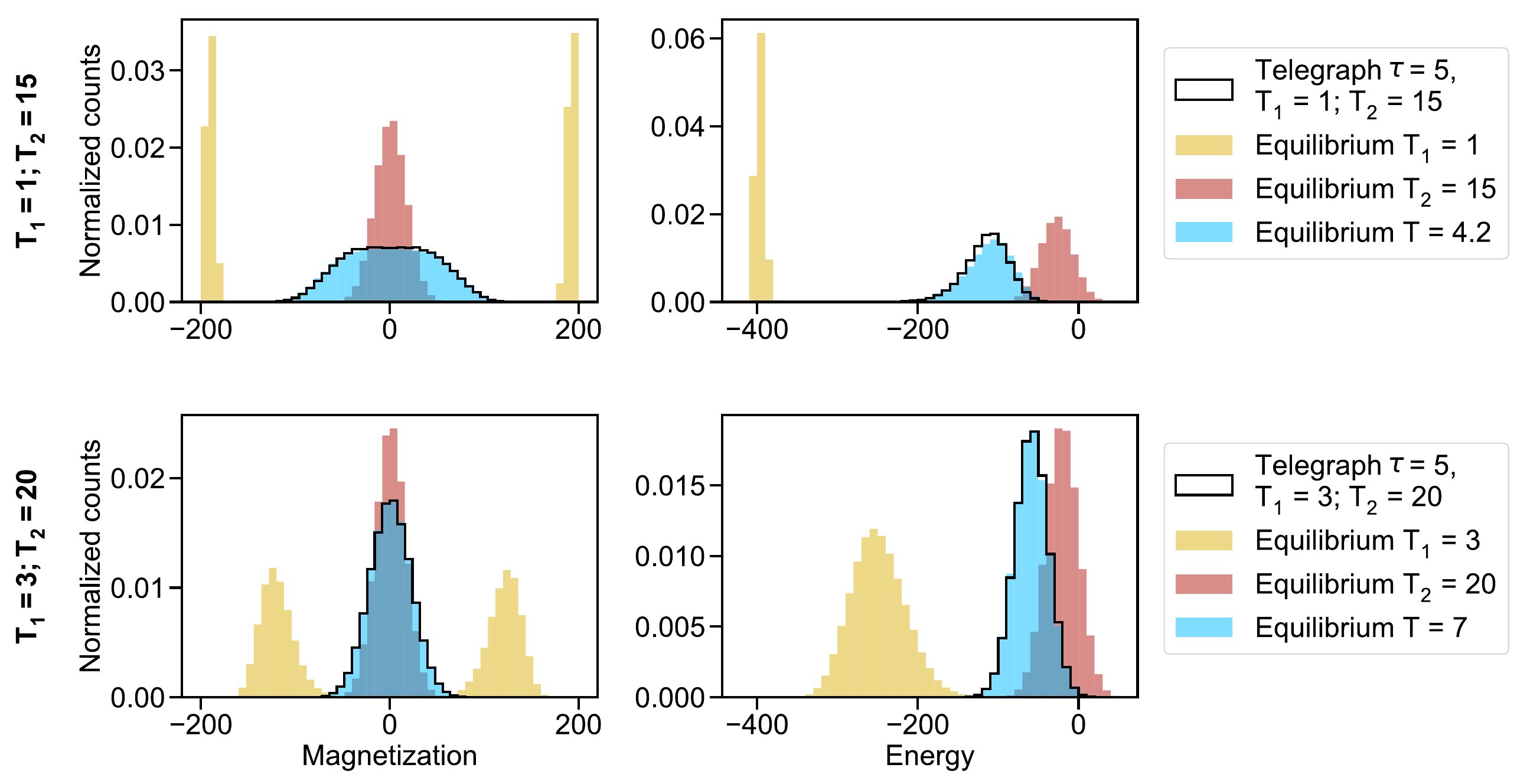}
	\caption{\textbf{Characterization of sequences subjected to fluctuating selection strength.} Histograms of the magnetization (left panels) and statistical energy (right panels) of each sequence are shown for sequences generated at equilibrium and for sequences subjected to fluctuating selection strength (i.e.\ sampling temperature), modeled by a telegraph process with timescale~$\tau$. All sequences are generated using the minimal model explained in Fig.~S1. In each case, 2048 sequences are generated. Those subjected to fluctuating selection strength are each generated starting from a different ancestor, but always using the same realization of the telegraph noise, with $\tau=5$, for a total duration such that 5000 mutations are accepted, using the same sequence length of 200 and the same Erd\H{o}s-Rényi random graph with probability 0.02 representing contacts as in Fig.~1. Two different pairs of temperatures $(T_1,T_2)$ are considered for the telegraph process (top and bottom panels). With the parameter values used in the top panels, the average TP fractions are  respectively equal to 0.9945 (standard deviation [std]:  0.0019) and 0.9997 (std:  0.0008) with the telegraph process (averaged over telegraph realisations) and at equilibrium with $T = 4.2$. With those used in the bottom panels, the average TP fractions are respectively equal to 0.9843 (std:  0.0028) and 0.9860 (std:  0.0051) with the telegraph process and at equilibrium with $T = 7$.
	}
	\label{fig:hist_erg_mag}
\end{figure}

\begin{figure}[htb!]
	\centering
	\includegraphics[width=0.8\columnwidth]{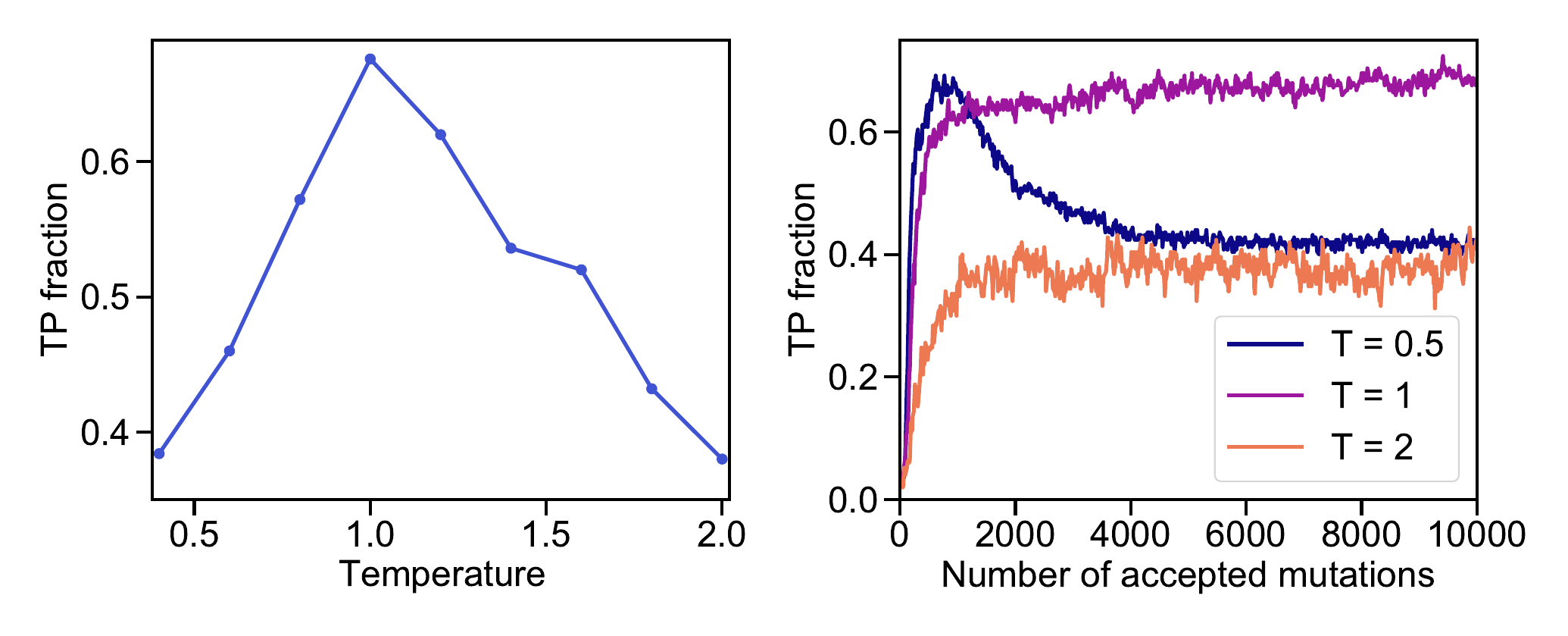}
	\caption{\textbf{Impact of sampling temperature on structural contact prediction: PF0004 family.} Left panel: As in Fig.~\ref{fig:tpfrac_vs_T}(a), the TP fraction for contact prediction is shown versus the sampling temperature, but for realistic data generated from a Potts model inferred on the PF0004 family. Right panel: The TP fraction for contact prediction is shown versus the number of accepted mutations, at different sampling temperatures, starting from random sequences. In both panels, in each case, we consider MSAs of 70,000 sequences generated from a Potts model inferred on a natural MSA of 39,277 sequences from the PF0004 family (with length 132), following Refs.~\cite{Lupo22,Figliuzzi18}. In the left panel, data obtained after 10000 accepted mutations is used at each temperature to infer contacts. This realistic data is thus similar to that considered in Fig.~2, except that here, we focus on data equilibrated at constant temperature, while in Fig.~2, we considered data generated under switching selection strength. As in Fig.~2, contact inference is performed using plmDCA~\cite{Ekeberg13,Ekeberg14}, with regularization strengths set to 0.01 and no phylogenetic reweighting. }
	\label{fig:tpfrac_vs_T_realistic}
\end{figure}

\begin{figure}[htb!]
	\centering
	\includegraphics[width=0.9\columnwidth]{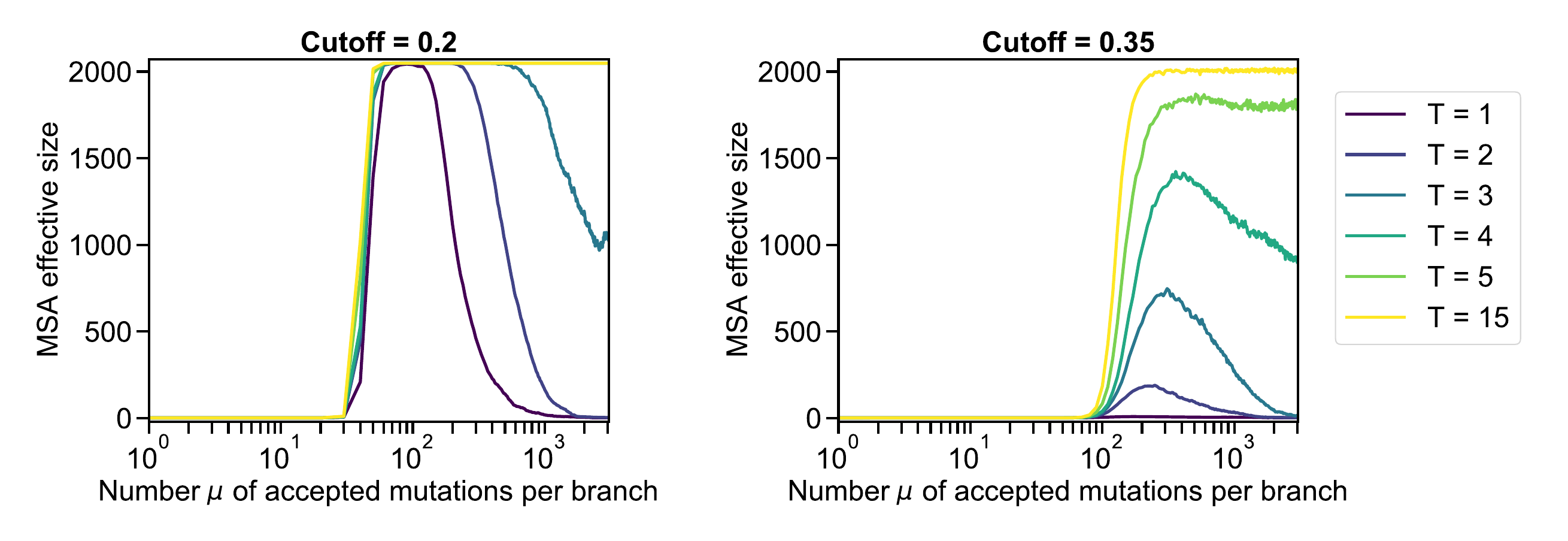}
	\caption{\textbf{Effective size of MSAs in the star phylogeny where selection is switched on.} We report the effective size of the MSA versus the number $\mu$ of mutations per branch for the datasets corresponding to Fig.~3, where we considered a star phylogeny starting from a random ancestral sequence. We compute the effective depth~\cite{Weigt09} of each MSA as $M_\mathrm{eff} = \sum_{i=1}^M w_i$, where for each sequence $i$, $w_i$ is the inverse of the number of sequences such that their Hamming distance to sequence $i$ is smaller than or equal to a cutoff value, which we set to $0.2$ or to $0.35$ in each panel. Recall that the actual number of sequences in each MSA is 2048, implying that $M_\mathrm{eff}\leq 2048$.}
	\label{fig:meff_vs_mu}
\end{figure}

\begin{figure}[htb!]
	\centering
	\includegraphics[width=0.95\columnwidth]{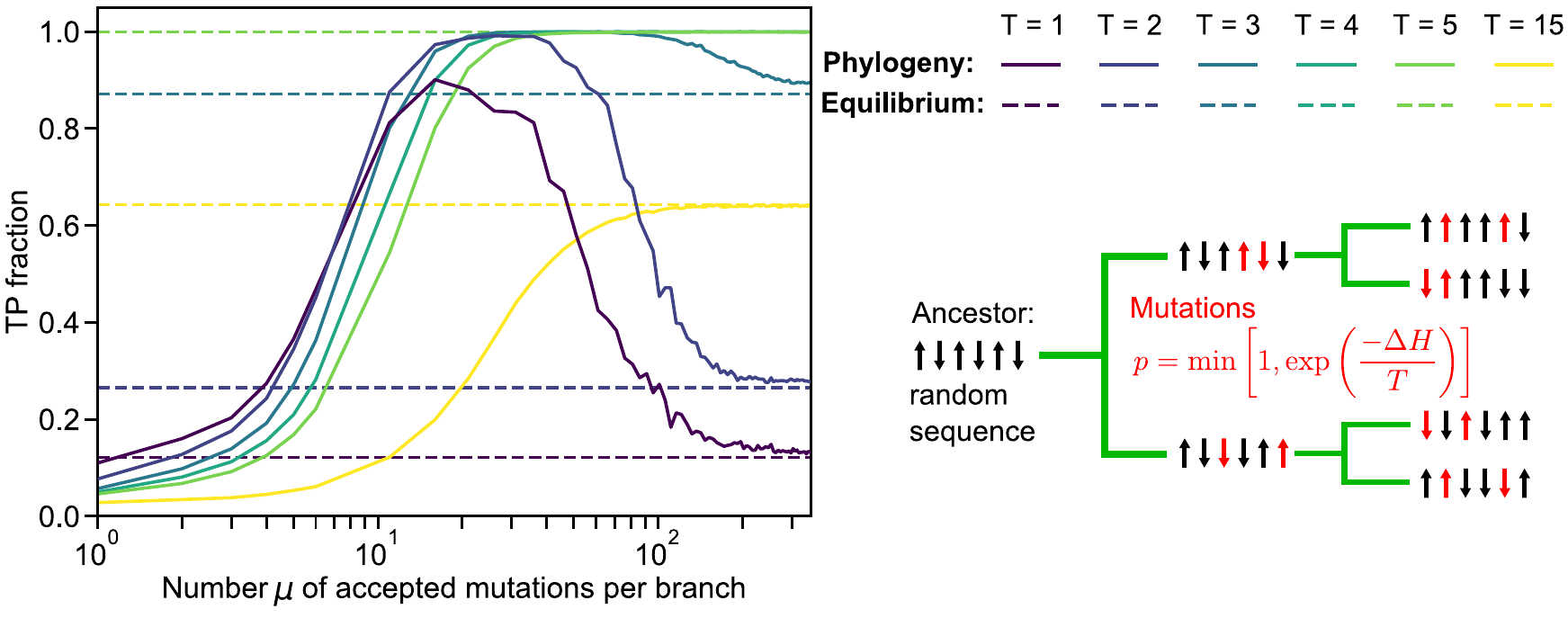}
	\caption{\textbf{Impact of switching selection on on contact inference using a binary branching tree.} 
		As in Fig.~3, the TP fraction is shown versus the number $\mu$ of accepted mutations per branch of a phylogenetic tree, starting from a random ancestral sequence. However, here, we use a binary branching tree (see schematic on the right) instead of a star phylogenetic tree. Sequences are generated as explained in Fig.~\ref{fig:fig_conceptual_outofeq}(a) and in the schematic on the right, using the same Erd\H{o}s-Rényi graph as in Fig.~1. 
		The phylogenetic tree comprises 11 generations of binary branching events, resulting in 2048 sequences at the leaves, which constitute our MSA, and $\mu$ mutations are accepted on each branch. We infer contacts via mfDCA~\cite{Marks11,Morcos11,Dietler2023} with pseudocount 0.01. Results are averaged over 100 replicates, each starting from a different random ancestor. Schematic (right): To minimally model the impact of selection being switched on, we take a random sequence as the ancestor, and we evolve it along a perfectly balanced binary branching tree (green) where mutations (red) are accepted with the probability in Eq.~(2), yielding selection for structural contacts. On each branch of the tree, Monte Carlo sampling is performed until $\mu$ mutations are accepted, with $\mu=2$ in our schematic. Hence, all branches in this simple tree have the same length.}
	\label{fig:tpfrac_vs_mu_binary}
\end{figure}

\end{document}